\documentclass[11pt]{article}
\usepackage{epsfig,amssymb,latexsym,color,amsmath,pifont}

\begin{document}

\twocolumn[

\title{\textcolor{blue}{\textbf{Chaotic dephasing in a double-slit scattering experiment}}}
\author{Zoran Levnaji\'c${}^{\; 1}$ and Toma\v z Prosen${}^{\; 1,2}$}
\date{}
\maketitle

\begin{flushleft}
${}^{1}$\textit{Department of Physics and Astronomy, University of Potsdam, Karl-Liebknecht-Street 24/25, Potsdam, Germany} \\
${}^{2}$\textit{Department of Physics, FMF, University of Ljubljana, Jadranska 19, Ljubljana, Slovenia}
\end{flushleft}

\begin{center} \begin{minipage}{6.5in}
We design a computational experiment in which a quantum particle tunnels into a billiard of variable shape and scatters out of it through a double-slit opening on the billiard's base. The interference patterns produced by the scattered probability currents for a range of energies are investigated in relation to the billiard's geometry which is connected to its classical integrability. Four billiards with hierarchical integrability levels are considered: integrable, pseudo-integrable, weak-mixing and strongly chaotic. In agreement with the earlier result by Casati and Prosen (Phys. Rev. A \textbf{72}, 032111 (2005)), we find the billiard's integrability to have a crucial influence on the properties of the interference patterns. In the integrable case most experiment outcomes are found to be consistent with the constructive interference occurring in the usual double-slit experiment. In contrast to this, non-integrable billiards typically display asymmetric interference patterns of smaller visibility characterized by weakly correlated wave function values at the two slits. Our findings indicate an intrinsic connection between the classical integrability and the quantum dephasing, responsible for the destruction of interference.
\end{minipage} \end{center}
]


\noindent \textbf{The chaotic phenomena in dynamical systems have been extensively studied and well understood over the last few decades. Yet, the quantum manifestations of classical chaos, despite largely investigated, are still not entirely clear. A possible bridge between the two is provided by the mechanism of quantum dynamical dephasing, occurring when a coherent quantum state undergoes a classically chaotic dynamics. In this work we report the results of a computational experiment that models scattering of a quantum particle through billiards whose classical dynamical properties depend on their shape. By relating the scattering outcomes with the integrability of the billiards (scatterers), we establish a clear connection between the classical chaos and the appearance of quantum dephasing. In particular, we find that a large dispersion of interference visibilities (contrasts) with respect to the energy sampling suggests a signature of classical chaos. Our results are in agreement with the previous computational and real experiments, and can themselves be obtained in a real laboratory experiment.}
\begin{center}
 ------------------------------------------------
\end{center}

\section{Introduction}   \label{Introduction}

The precise relationship between quantum and classical dynamics to this date still remains somewhat elusive, at both theoretical and experimental level. In particular, the understanding of quantum counterpart of macroscopic chaotic behavior is far from complete \cite{stockmann,haake}. As pointed out in \cite{casatiprosen}, there are two crucial properties of quantum motion that are in a direct contrast with common characteristics of classical chaos, namely: (\textit{i}) discrete energy-spectrum of bound quantum systems, rendering quantum dynamics quasi-periodic and hence decomposable in a finite Fourier-spectrum; (\textit{ii}) inherent stability of quantum dynamics, as a result of which initial errors always propagate only linearly with time.

Nevertheless, various general manifestations of classically chaotic phenomena in quantum dynamics have been widely studied over the last few decades \cite{stockmann,haake,ccgs,prosen}. Recently, some of the long-standing theoretical results in this direction received their experimental confirmation, such as the chaotic nature of quantum motion of the kicked top \cite{chaudhury}.

On the other hand, the phenomena of quantum decoherence provides a model to understand how the ma\-croscopic classical dynamics arises from the microscopic quantum dynamics \cite{zurek}. \textit{Dynamical decoherence}, as a specific form of quantum decoherence, was proposed in the context of studying the dynamical behavior of a quantum analog of a classically chaotic system \cite{casatiprosen,jacquod,fontezerbo}. In particular, it has been recognized that the classically chaotic dynamics is related to the mechanism of \textit{quantum dephasing}, responsible for the destruction of the phase-coherence of the initial wave function.

\begin{figure}[!ht]  \begin{center} 
       \includegraphics[height=5.cm,width=6.1cm]{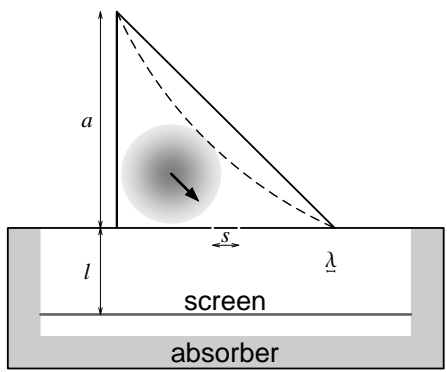}
  \caption{(from \cite{casatiprosen}) Time-dependent experiment: billiard with a variable geometry connected to the radiating domain through a double-slit opening. An initially coherent state freely evolves within the domain, 
           creating interference patterns on the screen.} \label{fig-cp-setup}
\end{center}  \end{figure}

\paragraph{Time-dependent experiment.} 
In their recent work \cite{casatiprosen}, Casati and Prosen designed a computational experiment in order to expose the mechanism of quantum dephasing arising due to (classically) chaotic dynamics. The time-dependent Schr\"odinger equation:
\begin{equation}     \dfrac{p^2}{2m} \Psi = i \hbar \dfrac{\partial \Psi}{\partial t}  \label{eq-timedep} \end{equation}
was solved numerically for a quantum particle moving freely inside the two-dimensional domain indicated by full line in Fig.\,\ref{fig-cp-setup}. The particle was initially prepared in a minimum uncertainty coherent state and placed inside the upper triangular region (billiard domain) with a given initial momentum.
Two small slits were connecting the billiard domain with the lower radiating domain, allowing the particle to leak out in form of two interfering probability currents during the time-evolution. The particle's wave number $k$ defining the de Broglie wavelength $\lambda = 2\pi/k$ was chosen in a way to have $\lambda$ much bigger than the width of the slits, and much smaller than the distance between them. The interference patterns, created by time-integrating the perpendicular component of the probability current on the horizontal screen below, were observed. The experiment went on until the probability to find the particle within the billiard domain was vanishingly small.

\begin{figure}[!ht] \begin{center}
       \includegraphics[height=5.5cm,width=7.5cm]{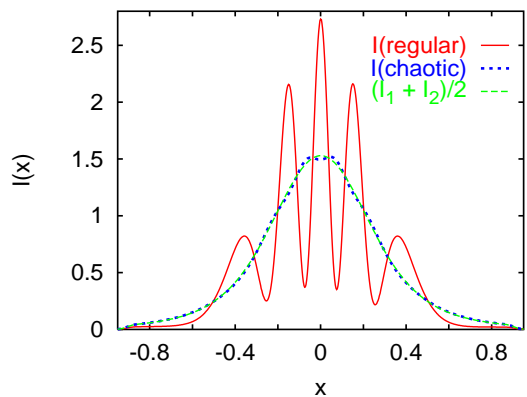} 
  \caption{(from \cite{casatiprosen}, color online) Results of the time-dependent experiment: while the integrated interference pattern for the case of integrable billiard shows constructive double-slit fringes,  
           the pattern displayed in the chaotic case appears to be identical to the superposition of two uncorrelated sources.} \label{fig-cp-result}
\end{center} \end{figure}

The experiment was repeated for two different sha\-pes of billiard domains: the right-angle isosceles triangle, and the Sinai-type (defocusing) billiard obtained by replacing this triangle's hypotenuse with a circular arc (dashed line in Fig.\,\ref{fig-cp-setup}). These two billiards have drastically different classical dynamical properties: while the former generates regular integrable motion, the latter is known to be a strongly chaotic K-system (strong mixing) \cite{cfs}. Fundamentally different results were found for the integrated interference patterns, as shown in Fig.\,\ref{fig-cp-result}. In the integrable case the usual double-slit fringes characterizing constructive interference were recovered, whereas in the chaotic case the fringes were entirely absent and only a bell-shaped unimodal intensity distribution was found. 
In particular, the chaotic billiard's profile was essentially identical to the situation of two slits emitting independent probability currents. While the classically regular/integrable dynamics preserved the initial phase-coherence of the particle, the chaotic dynamics produced pseudo-random phase shifting, ultimately generating ``uncorrelated'' output currents. Classically chaotic time-evolution induced the quantum dephasing process, responsible for the destruction of particle's phase-coherence, and thus causing the disappearance of the expected double-slit fringes. \\[0.1cm]

This result has been recently confirmed in a real experiment investigating wave chaos \cite{tang-water}. Furthermore, similar computational experiments have confirmed an intrinsic relationship between classical chaos and quantum dephasing in a wide range of examples \cite{jacquod}, while the important role of the geometric symmetry has been pointed out in \cite{fontezerbo}.

In this work we present the results of a new computational experiment, constructed along the similar lines as the time-dependent experiment described above, but now considering the energy domain instead of the time domain. We consider stationary scattering of quantum particles with different energies through the billiards of variable geometries, designed to include a double-slit opening. The interference patterns are observed on the screen over a wide energy-interval, and statistically investigated in relation to the billiards' geometries. As we show, the findings from time-dependent experiment fully extend to the scattering case, shedding new light on the mechanism of dynamical dephasing. Moreover, two new billiards with the particular non-integrable geometries, whose dynamics interpolates between integrable and strongly chaotic, are added in order to study the appearance of dephasing in relation to the ergodic hierarchy of dynamical systems. In opposition to \cite{fontezerbo}, we avoid the presence of any kind of geometric symmetries of our billiards, and focus exclusively on their integrability properties.

For each energy value we obtain an interference pattern whose properties are essentially given by the difference and ratio of the wave function's phases and magnitudes, respectively, at the positions of the two slits. We specifically consider the distribution of the interference visibilities, given as the range of ratios of the wave function intensities. Our main finding is that in the case of chaotic classical dynamics, this distribution is much wider than in the case of integrable dynamics: a ``typical'' interference pattern produced by each of the non-integrable billiards is much more likely have lower visibility than the one created by the integrable billiard.

This paper is organized as follows: in the next Section we describe in detail the design of our computational experiment and the parameters used. In Section \ref{Results} we expose the results found, and finally provide a discussion and conclusions in Section \ref{DiscussionConclusions}.


\section{Scattering experiment} \label{Scatteringexperiment}

In this Section we explain the construction of our scattering apparatus, provide the parameter values used in the experiment, and outline the computational details.

\begin{figure*}[!ht] \begin{center}
       \includegraphics[height=10.cm,width=11.8cm]{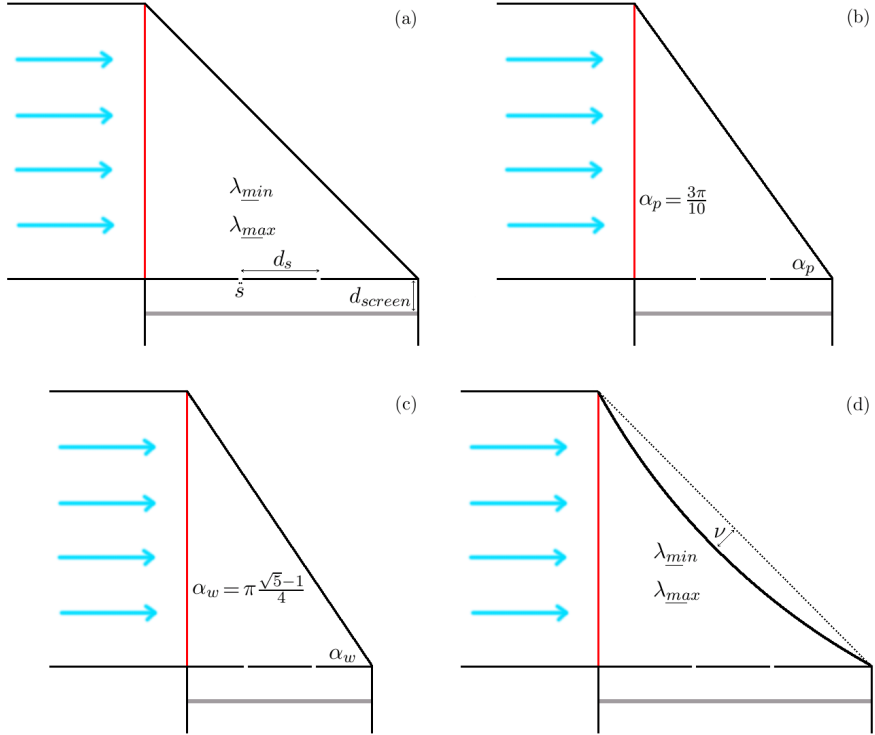}
  \caption{(color online) Scattering experiment set-up involving four billiards with different classical dynamical properties: (a) integrable, (b) pseudo-integrable, (c) weak-mixing, and (d) chaotic. 
    Billiards are attached to an input channel of unitary width and separated from it by a potential tunneling barrier that replaces that billiard's side. Base of each billiard contains a double-slit opening, leading to the 
    output channel where the interference patterns are recorded on the screen at a distance $d_{screen}$. Particle's wavelengths are shown for a comparison with slits width $s$, distance between the slits $d_s$, and width of 
    the circular segment $\nu$ (for chaotic billiard in (d)). Arrows illustrate the momentum of the incoming particles (plane waves) that impact orthogonally on the potential barrier. We take the plane wave direction to be $x$-coordinate, and the direction of the scattered currents to be $y$-coordinate. In all the figures to follow, unless noted otherwise, we use this scheme for representing the results for four billiards.} \label{fig-s-setup}
\end{center} \end{figure*}

We consider a isosceles right triangle billiard with the double-slit opening on its base (as the one used in time-dependent experiment), shown in Fig.\,\ref{fig-s-setup}a. We replace its left side with a potential tunneling barrier of height $V_{barrier}$ shown in red, and use it to connect the billiard to the \textit{input channel} having the width equal to that triangle's side. All other billiard borders shown in black are assumed to be infinite potential barriers. The billiard's base containing the slits is connected to the \textit{output channel} having its width equal to it. Inside the output channel at a distance $d_{screen}$ from the base, we put the screen where the interference patterns are observed and recorded. Thus constructed scattering system consists of three parts: input channel used to inject the particle into the triangular billiard (illustrated as cyan arrows), the billiard itself used as the scatterer, and the output channel where the interference patterns are observed.
The scattering experiment is done for four different billiards, out of which we have just described the first. Other billiards are designed by modifying this basic triangular billiard, and named through their classical dynamical properties. In particular:
\begin{itemize}
\item \textit{integrable} -- (just described) isosceles right-angle triangle with angles $\frac{\pi}{4},\frac{\pi}{4},\frac{\pi}{2}$, shown in Fig.\,\ref{fig-s-setup}a. Classically, it generates exclusively regular, periodic or quasi-periodic motion.
\item \textit{pseudo-integrable} -- right-angle triangle with angles $\frac{2\pi}{10},\frac{3\pi}{10},\frac{\pi}{2}$, classically a pseudo-integrable system \cite{gutkin}, shown in Fig.\,\ref{fig-s-setup}b.
\item \textit{weak-mixing} -- right-angle triangle with two angles irrational with $\pi$, classically conjectured to be a weak-mixing system \cite{weak-conjecture}, shown in Fig.\,\ref{fig-s-setup}c. For angles we 
choose $\pi \frac{3-\sqrt{5}}{4},\pi \frac{\sqrt{5}-1}{4},\frac{\pi}{2}$. 
\item \textit{chaotic} -- the billiard obtained by replacing the triangle's hypotenouse with a circular arc of radius $R$, as done in the time-dependent experiment, known to be a strongly chaotic K-system \cite{cfs}, shown in Fig.\,\ref{fig-s-setup}d.
\end{itemize}
A quantum particle in the form of a plane wave with definite energy $E = p^2/2$ is injected into the billiard through the input channel by tunneling the potential barrier, and is then scattered out through the slits. We model this process by numerically solving the ``on-shell'' time-independent scattering problem for the two-di\-men\-sional Schr\"odinger equation
\begin{equation} -\frac{\hbar^2}{2m} \left( \frac{\partial^2}{\partial x^2} + \frac{\partial^2}{\partial y^2} \right) \Psi = E \Psi \label{eq-scattering}  \end{equation}
with scattering boundary conditions. The scattering wave function $\Psi$ consists of three spatially disjoint parts:
\begin{itemize}
\item  $\Psi_1$ in the input channel -- incoming plane wave  
\item  $\Psi_2$ inside the billiard -- scattering state 
\item  $\Psi_3$ in the output channel -- scattered waves
\end{itemize}
The final solution $\Psi = \Psi_1 \oplus \Psi_2 \oplus \Psi_3$ is obtained by imposing the spatial continuity condition between $\Psi_1$ and $\Psi_2$, and between $\Psi_2$ and $\Psi_3$, along with the boundary conditions given by the billiard's geometry (infinite potential barriers) and the incoming particle. While using $\Psi_1$ as input, we will investigate the properties of $\Psi_2$, and in particular focus on $\Psi_3$ which generates the interference patterns related to the billiard's integrability properties and the particle's energy $E$.

Majority of the probability current is immediately reflected back from the potential barrier and does not undergo the scattering process. However, if penetrates the tunneling barrier, the particle is much more likely to exit the billiard through the slits than back through the input channel (penetrating the barrier again). The presence of the barrier drastically reduces the intensity of the probability currents exiting the billiard and interfering, but on the other hand allows the penetrating particle to ``feel'' the billiard's geometry much better. The barrier's height is kept equal for all billiards. In opposition to the time-dependent experiment, we are now not integrating the recorded interference patterns (in energy), but instead examine their properties statistically over the entire energy-interval. Also, we do not use any absorbing potential at the edges of the output channel. We took the input channel to have the maximum possible width (that of the billiard), in order to obtain the best focus  in particle's direction (momentum) while allowing for maximal indeterminacy in position. This way we optimize the directionality of the plane wave's impact on the barrier, assuring that the scattering outcome as best possible reports about the properties of the billiard. Besides the barrier, the strength of probability currents also depends on the eigen-energies of the closed billiard, that act as resonances in the scattering process. The transport through the scattering apparatus is enhanced when the particle's energy corresponds to one of the scatterer's (billiard's) resonant energies. However, we will consider an ensemble of all the recorded interference patterns with equal statistical weights, regardless of the particles' energies.

By considering the four billiards we investigate the scattering process in relation to a hierarchy of systems having different classical ``chaoticity'' levels. Thus, we obtain a more detailed relationship between the classical dynamical properties and the appearance/strength of the quantum dephasing. For all billiards we set the width of the input channel (and hence the length of the billiard's potential barrier side) to be 1, therefore fixing the billiards' areas. This implies that the billiards will have slightly different areas, while pseudo-integrable and weak-mixing billiards will also have their base side somewhat shorter (and hence their output channels somewhat narrower). The chaotic billiard is designed with a fixed circular arc replacing the hypotenouse, and with the base equal to 1 (that is to say, by cutting a circular segment away from the isosceles triangle billiard).

\paragraph{The parameter values used.} We consider the system of units such that not only the width of the billiard, but also the Planck's constant and the particle's mass equal unity, $\hbar=m=1$. We seek to have slits width $s$ much smaller than the particle's wavelength $\lambda$, which itself has to be much smaller than the distance between the slits $d_s$:
\begin{equation}     s \ll \lambda \ll d_s   \label{eq-slambdads}    \end{equation}
Furthermore, for the chaotic billiard we take the width of the cut circular segment $\nu$ to be larger (but comparable) to the particle's wavelength $\lambda$ (cf. Fig.\,\ref{fig-s-setup}d):
\[ \lambda  \lesssim \nu \]
in order for ``chaoticity'' of this billiard to be ``visible'' by the quantum particle. Accordingly, we set the parameters and dimensions used in our experiments as follows: \\[0.1cm]
$\ast$ the lengths of the billiards' sides -- integrable: $1,1,1.414$; pseudo-integrable: $1,0.727,1.236$; weak-mixing: $1,0.684,$ $1.212$; chaotic: $1,1,1.428$. \\[0.1cm]
$\ast$ the billiard's areas $A$ -- integrable: $0.5$; pseudo-integrable: $0.363$; weak-mixing: $0.342$; chaotic: $0.414$. \\[0.1cm]
$\ast$ curvature radius for chaotic billiard: $R=2.786$, the width of the cut circular segment: $\nu=0.091$. \\[0.1cm]
$\ast$ width of the slits $s$ -- integrable and chaotic: 0.017, pseudo-integrable and weak-mixing: 0.016. \\[0.1cm]
$\ast$ distance between the slits $d_s$ -- integrable and chaotic: 0.287, pseudo-integrable and weak-mixing:  0.257. \\[0.1cm]
$\ast$ the energy-interval of incoming particles was measured in terms of particle's wave number $k=\sqrt{2E}$, and set from $k_{min}=85$ ($\lambda_{max}=0.074, E_{min}=3612$), to $k_{max}=94$ ($\lambda_{min}=0.067, E_{max}=4418$). This choice optimizes the condition Eq.\,(\ref{eq-slambdads}), along with the condition $\lambda  \lesssim \nu$ stated above. \\[0.1cm]
$\ast$ the $k$-step between two particles of consecutive $k$-values: $\delta k = 0.001$. We made 9000 experiments for each billiard, with particles uniformly spaced in $k$, covering the interval of $k$ mentioned above. \\[0.1cm]
$\ast$ the height of the potential tunneling barrier: $V_{barrier}=7200$, corresponding to $k_{barrier}=\sqrt{2V_{barrier}}=120$. This value was set by maximizing the dephasing effects while still having measurable probability currents (see Section \ref{DiscussionConclusions}). The barrier's energy is thus roughly twice bigger than average particles' energies. \\[0.1cm]
$\ast$ distance from the billiard's base at which the interference patterns are observed: $d_{screen} =0.12$. \\[0.3cm]
Therefore, in our system of units the Eq.\,(\ref{eq-scattering}) reduces to: 
\begin{equation} \left( \frac{\partial^2}{\partial x^2} + \frac{\partial^2}{\partial y^2} + k^2 \right) \Psi = 0 \label{eq-scattering-ourunits}  \end{equation}
We numerically solved this equation (with appropriate boundary conditions) for all $k$-values and parameters as described above, and for all four billiards.

\paragraph{The numerical implementation.} The Eq.\,(\ref{eq-scattering-ourunits}) is solved numerically by discretizing the space where the wave function $\Psi$ is defined. The space within the billiards is covered by a grid of $\frac{B(B+1)}{2}$ mesh points, the potential barrier, input and output channels are each covered by a sequence of $B$ points, while each of the slits is covered by $S$ points. Each point is labeled in two dimensions as $\psi_{n,m}$, and represents the wave function complex value at its coordinates. The Laplacian is discretized as follows: for a given point $\psi_{n,m}$ (e.g. inside the billiard), the Eq.\,(\ref{eq-scattering-ourunits}) reads:
\begin{equation}  \begin{array}{rrr}
\Delta_x^{-2} \cdot (\psi_{n+1,m} +\psi_{n-1,m}-2\psi_{n,m}) & + & \\[0.1cm]
\Delta_y^{-2} \cdot (\psi_{n,m+1} +\psi_{n,m-1}-2\psi_{n,m}) & + & \\[0.1cm]
 k^2\psi_{n,m} &=& 0 
\end{array}   \label{eq-scattering-discretization}  \end{equation}
where $\Delta_x$ and $\Delta_y$ are the space discretization steps in horizontal and vertical direction. 
This way we construct a linear system of dimensionality $N = \frac{B(B+1)}{2} + 3B + 2S$ in the form 
\begin{equation}  {\mathcal A} \bar{\psi} = \bar{\eta_0}  \label{eq-scattering-linear}  \end{equation}
where the $N$-dimensional vector $\bar{\psi}$ is the solution of the scattering problem containing the points $\psi_{n,m}$, vector $\bar{\eta_0}$ contains the inputs defining the incoming plane wave (and zeros for other points as in example of Eq.\,(\ref{eq-scattering-discretization})), while the $N\times N$ matrix ${\mathcal A}$ contains discretized two-dimensional Laplacian and the conversion coefficients for other parts of the scattering apparatus. The vector $\bar{\psi}$ includes $B$ values defining the wave function inside the output channel, used to compute the interference pattern.

The grid points for the integrable billiard form a regular square lattice $\Delta_x=\Delta_y$. For other billiards we construct the appropriate grids in accordance with their geometries, so that all the boundary grid points lie exactly on the geometrical billiard's boundary. For the chaotic billiard we use the same `trick' as in the time-dependent experiment: the spacing between the grid's horizontal rows $\Delta_y$ is made non-uniform in order to produce a smooth curvature modeling the circular arc that replaces the hypotenouse. For pseudo-integrable and weak-mixing billiards, the spacing between the grid's vertical columns $\Delta_x$ is reduced by a factor of $\tan \frac{3\pi}{10}$ and $\tan \pi\frac{\sqrt{5}-1}{4}$ respectively, in order to model their geometries. In our experiments, we used the values of $B=180$ for all billiards (whereas we have carefully checked that numerics is robust against variation of $B$ in this range), along with $S=3$ for integrable and chaotic, and $S=4$ for pseudo-integrable and weak-mixing billiards (this explains slightly different values of slits distance and width for these billiards, as described in the previous paragraph).

The total numerical error in the experiments was estimated by computing the sum of the probability flux through the slits $\phi_{out}$ and the probability flux back-scattered through the input channel (barrier) $\phi_{back}$, and comparing this sum with the probability flux of the incoming wave -- normalized to $1$. The total unitarity error given by $|(\phi_{out}+\phi_{back}) - 1|$ was typically of the order of $10^{-5}$, and only in the vicinity of billiards' resonances went up to $10^{-2}$.


\section{Results} \label{Results}

In this Section we expose the results of the computational experiments designed and carried out as described in the previous Section. We systematically compare the findings among four billiards.

\begin{figure*}[!ht] \begin{center} 
       \includegraphics[height=10.cm,width=11.8cm]{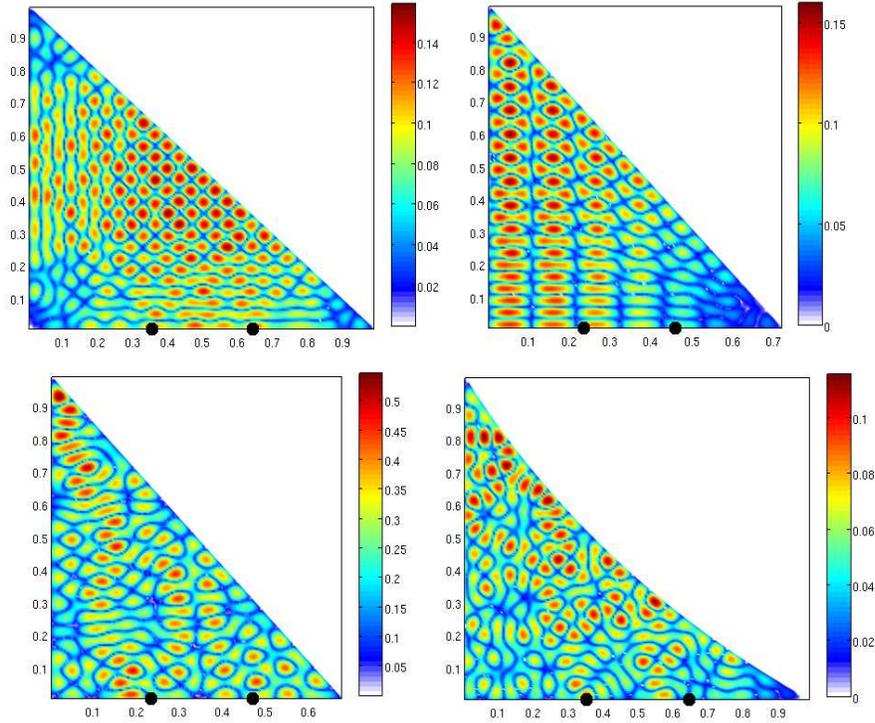}
  \caption{(color online) Densities of the scattering states $|\Psi_2|^2$ inside the billiards for the $k$-value of $k=90$. The scale of $|\Psi_2|^2$ is shown on the right of each billiard.
                          Big dots indicate the positions of the slits. Plots correspond to billiards as depicted in Fig.\,\ref{fig-s-setup}.} \label{fig-scatteringstates}
\end{center}  \end{figure*}

\paragraph{The scattering states and resonance profiles.} We examine the billiard's scattering states given by the wave function $\Psi_2$ defined within the billiards. In Fig.\,\ref{fig-scatteringstates} we show the color-plot of the densities $|\Psi_2|^2$ for all billiards for the wave number $k=90$. The wave function density for the integrable billiard is clearly very regular, while the chaotic billiard's density appears to be rather irregular. Pseudo-integrable billiard displays a density structurally similar to the integrable case, while the weak-mixing density seems closer to the chaotic one.
The regularity of the wave function densities neatly follows the hierarchy of the billiards' classical integrability levels. The scattering wave function's properties reflect the billiards' classical dynamical properties. The characteristic lengths of densities from Fig.\,\ref{fig-scatteringstates} correspond to the incoming particle's wavelength. They do not vary substantially among the billiards, as $k=2\pi/\lambda$ is the same for all billiards shown in Fig.\,\ref{fig-scatteringstates}. The densities for different wave numbers $k$ qualitatively preserve the regularity/irregularity of their structural patterns.

\begin{figure*}[!ht] \begin{center}
       \includegraphics[height=9.cm,width=14.cm]{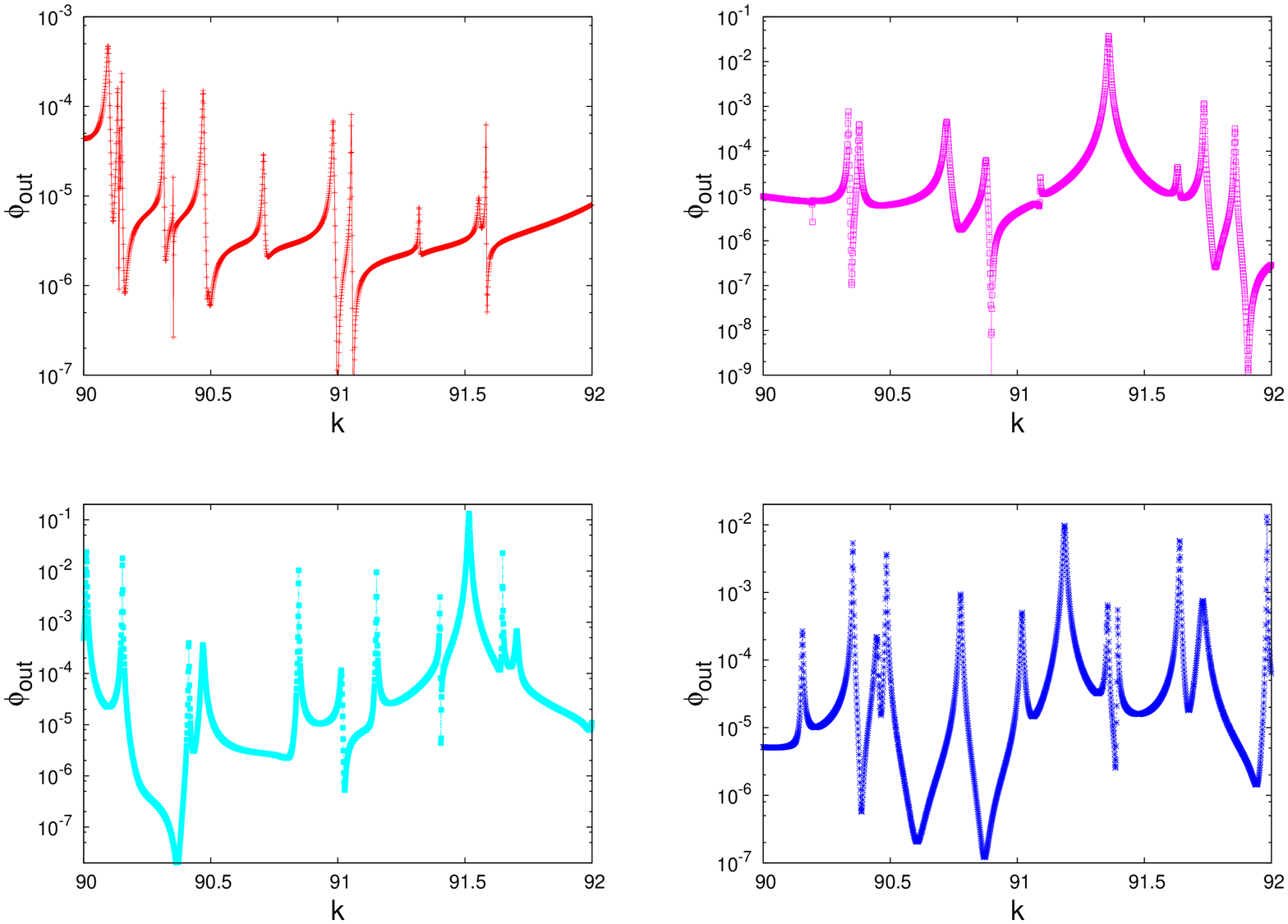} 
  \caption{Profiles of $\phi_{out}$-value as s function of incoming particle's wave number $k$ for all the four billiards, for the wave number interval from $k=90$ to 
      $k=92$. Log-scale is used in $\phi_{out}$ display more clearly the structure of the resonances. Profiles correspond to billiards as depicted in
      Fig.\,\ref{fig-s-setup}.} \label{fig-resonances}
\end{center} \end{figure*}

We define the probability flux leaking through the slits $\phi_{out}$ as the total probability current passing through any perpendicular cross-section ${\mathcal C}_{out}$ of the output channel (such as the screen for instance):
\[  \phi_{out} = -\mbox{Im} \; \int_{{\mathcal C}_{out}} \Psi^*_3 \frac{\partial}{\partial y} \Psi_3 \] 
which sensitively depends on the particle's energy, and its relationship with the billiard's resonant eigen-ener\-gies. We investigate the values of the flux $\phi_{out}$ as a function of the particle's wave number $k$, and report our findings in the Fig.\,\ref{fig-resonances}. For all the four billiards the profile of $\phi_{out}$-values displays peaks indicating the positions of billiards' scattering resonances (roughly corresponding to the eigen-energies of the closed billiards).
The density of peaks slightly varies among the billiards as a consequence of their different areas. This density $\frac{dN_r}{dk}$ is proportional to the billiard's area $A$, which is known as the Weyl's law \cite{weyl}: 
\[ \frac{dN_r}{dk} = \frac{A}{2\pi} k  \]
(for this reason we used particles equally spaced in wave number rather than in energy). Since the wave number interval is the same for all billiards in Fig.\,\ref{fig-resonances}, the ratio of number of peaks roughly follows the ratio of billiards' areas, i.e. $0.5:0.363:0.342:0.414$. From this formula one can also estimate the total number of peaks in the given $k$-interval. Note the different shapes of peaks: in the integrable case, immediately after the resonant $k$-value, $\phi_{out}$ sharply decays, whereas for other billiards the resonant peaks are relatively symmetric. Also note that the densities of Fig.\,\ref{fig-scatteringstates} evaluated for $k=90$ are on different scales (weak-mixing case in particular). This is a consequence of the particle with $k=90$ producing different scattering outputs for different billiards: as clear from Fig.\,\ref{fig-resonances}, for $k=90$ the $\phi_{out}$-value is biggest in the weak-mixing case.

\paragraph{The interference patterns.} In the reminder of this Section, we analyze in detail the interference properties of the scattered probability currents, comparing them among the billiards. The intensity $I(x)$ of the vertical component of the probability current within the output channel, computed at the distance $d_{screen}$ from the billiard's base is:
\begin{equation} I(x) = -\mbox{Im} \left( \Psi_3^\ast \frac{\partial}{\partial y} \Psi_3 \right) \; \Bigg|_{d_{screen}}   \label{eq-interference}  \end{equation}
where $x$ is the horizontal coordinate parameterizing the screen. For an interference pattern $\Sigma (x)$ we intend the normalized intensity $I(x)$:
\[ \Sigma (x) = \dfrac{I(x)}{\int_{\mbox{\scriptsize screen}} I(x) dx}  \]
This way we eliminate the pattern's dependence of the total output flux $\phi_{out} (k)$ and focus only on its interference properties, such as symmetry and visibility (contrast ratio of intensity between adjacent minima and maxima). Because of being a result of the interference between two probability currents, the patterns' properties are linked to the wave function's values at the centers of the (almost point-like) slits $\Psi_{s1}$ and $\Psi_{s2}$ (connecting $\Psi_2$ and $\Psi_3$). Namely, we can approximately imagine the field $\Psi_3$ to be a result of a simple interference between two point sources, $\Psi_{s1}$ and $\Psi_{s2}$. In order to examine the patterns $\Sigma (x)$ systematically, we introduce the quantities $\Delta_\rho$ and $\Delta_\phi$ as follows:
\begin{equation} \begin{array}{lll}
\Delta_\rho &=& 2 \; \mbox{Re} \left(\ln \dfrac{\Psi_{s1}}{\Psi_{s2}} \right) = \ln  \dfrac{|\Psi_{s1}|^{2}}{|\Psi_{s2}|^{2}} \\[0.4cm]
\Delta_\phi &=& \mbox{Im} \left(\ln \dfrac{\Psi_{s1}}{\Psi_{s2}} \right) \; =  \arg \Psi_{s1} - \arg \Psi_{s2} 
\label{eq-pd-definition} \end{array} \end{equation} 
$\Delta_\rho$ reports the ratio of the wave function densities at the slits. Thus, $\Delta_\rho=0$ means that the particle leaks through both slits with equal intensities, which is directly related to the interference visibility. $\Delta_\rho=0$ implies the maximal visibility and hence the infinite contrast ratio in the far field (i.e. perfectly {\em constructive} interference corresponding to the usual text-book double-slit experiment). For increasing $|\Delta_\rho|$ visibility gradually worsens, and for $|\Delta_\rho| \gg 1$ the interference pattern is practically washed away. $\Delta_\phi$ gives the value of the phase-shift between the wave function values at the slits and directly controls the (a)symmetry of the interference pattern. $\Delta_\phi=0$ indicates a perfectly symmetric $\Sigma$. Numerically, we compute the $\Psi_{s1,s2}$ by averaging the values for all the space-discretization points inside the slits (slits are very small anyway due to $s \ll \lambda$). 

\begin{figure*}[!ht] \begin{center}
       \includegraphics[height=9.cm,width=12.5cm]{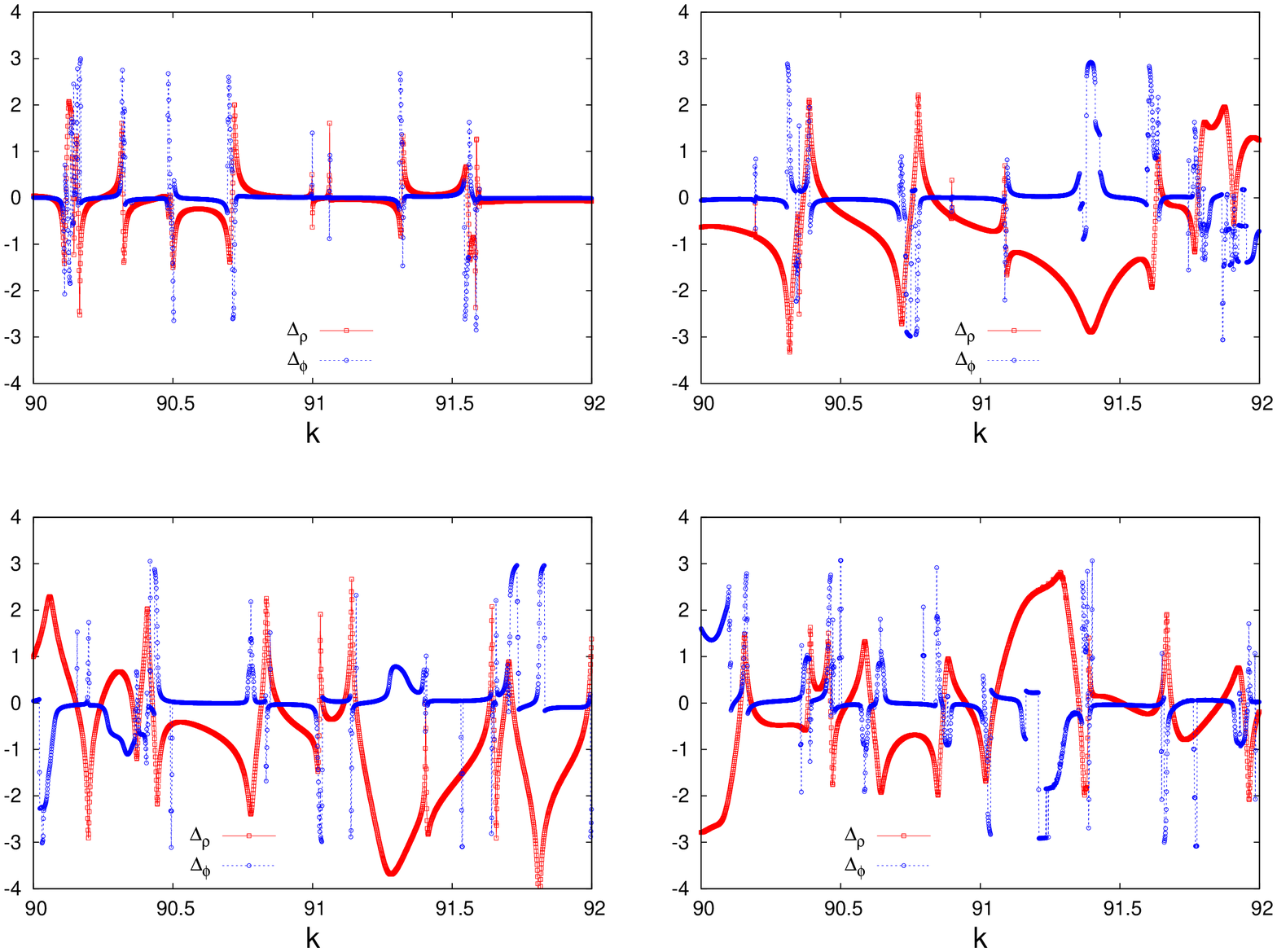}
  \caption{(color online) Profiles of $\Delta_\rho$-values in red and $\Delta_\phi$-values in blue (cf. Eq.\,(\ref{eq-pd-definition})) as functions of incoming particle's wave number for all the four billiards, 
           for the $k$-interval from $k=90$ to $k=92$. Profiles correspond to billiards as in Fig.\,\ref{fig-s-setup}.} \label{fig-pd}
\end{center}  \end{figure*}

\begin{figure*}[!ht] \begin{center} 
       \includegraphics[height=10.cm,width=12.4cm]{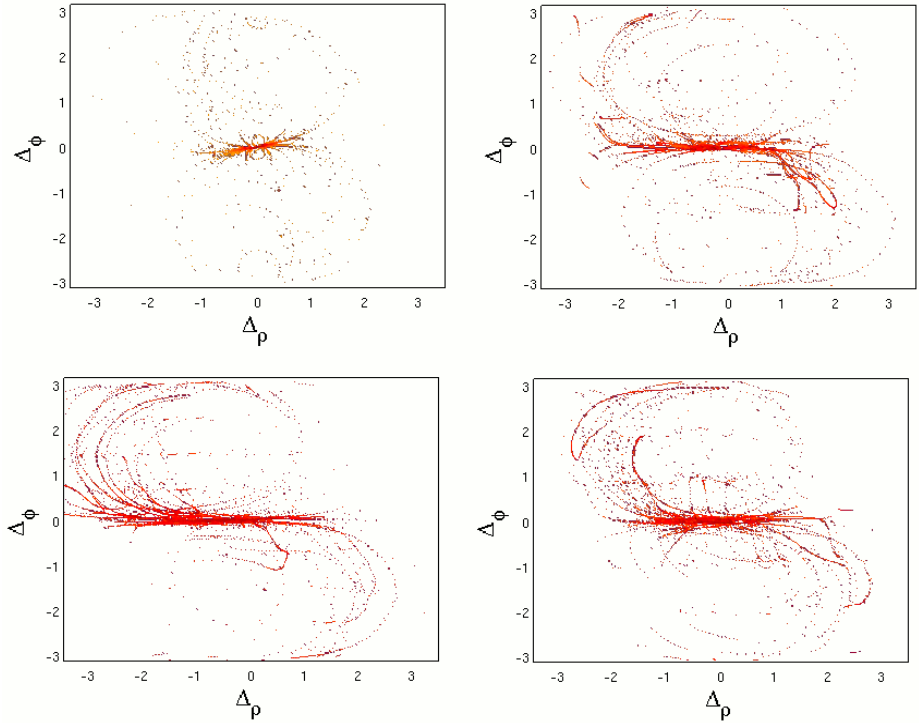}
  \caption{(color online) Two-dimensional distributions of $(\Delta_\rho,\Delta_\phi)$-values obtained for all experiments done in the $k$-interval from $k=85$ to $k=94$. The fraction of the 
      interference patterns having a certain $(\Delta_\rho,\Delta_\phi)$-value is shown (in log-scale). Plots correspond to billiards in order as depicted in Fig.\,\ref{fig-s-setup}.} \label{fig-pdscatterplot}
\end{center}  \end{figure*}

In Fig.\,\ref{fig-pd} we show the profiles of $\Delta_\rho$ and $\Delta_\phi$ as functions of the wave number $k$. 
Note a very different behaviors for different billiards. The integrable case exhibits a smooth behavior with mostly $\Delta_\rho=\Delta_\phi=0$, in addition to sporadic jumps overlapping with the resonances. 
In contrast to this, the non-integrable billiards display a much more dispersed fluctuating behaviors that intensify with the billiards' ``chaoticity'' level. In particular, the $\Delta_\rho$ profile (although relatively smooth) shows large variations, mostly staying far from $\Delta_\rho=0$ even for $k$-values away from the resonances. In the integrable case, the resonances are very narrow and the transport through the cavity is very small for most values of $k$. This in particular corresponds to the regions where $\Delta_\rho$ and $\Delta_\phi$ are very small. Nevertheless, the variations of the interference visibility $\Delta_\rho$ are smaller in the integrable case than in the non-integrable cases even in the resonance regions (see also Fig.\,\ref{fig-pdscatterplot}).

\begin{figure*}[!ht] \begin{center}
  \includegraphics[height=7.5cm,width=16.cm]{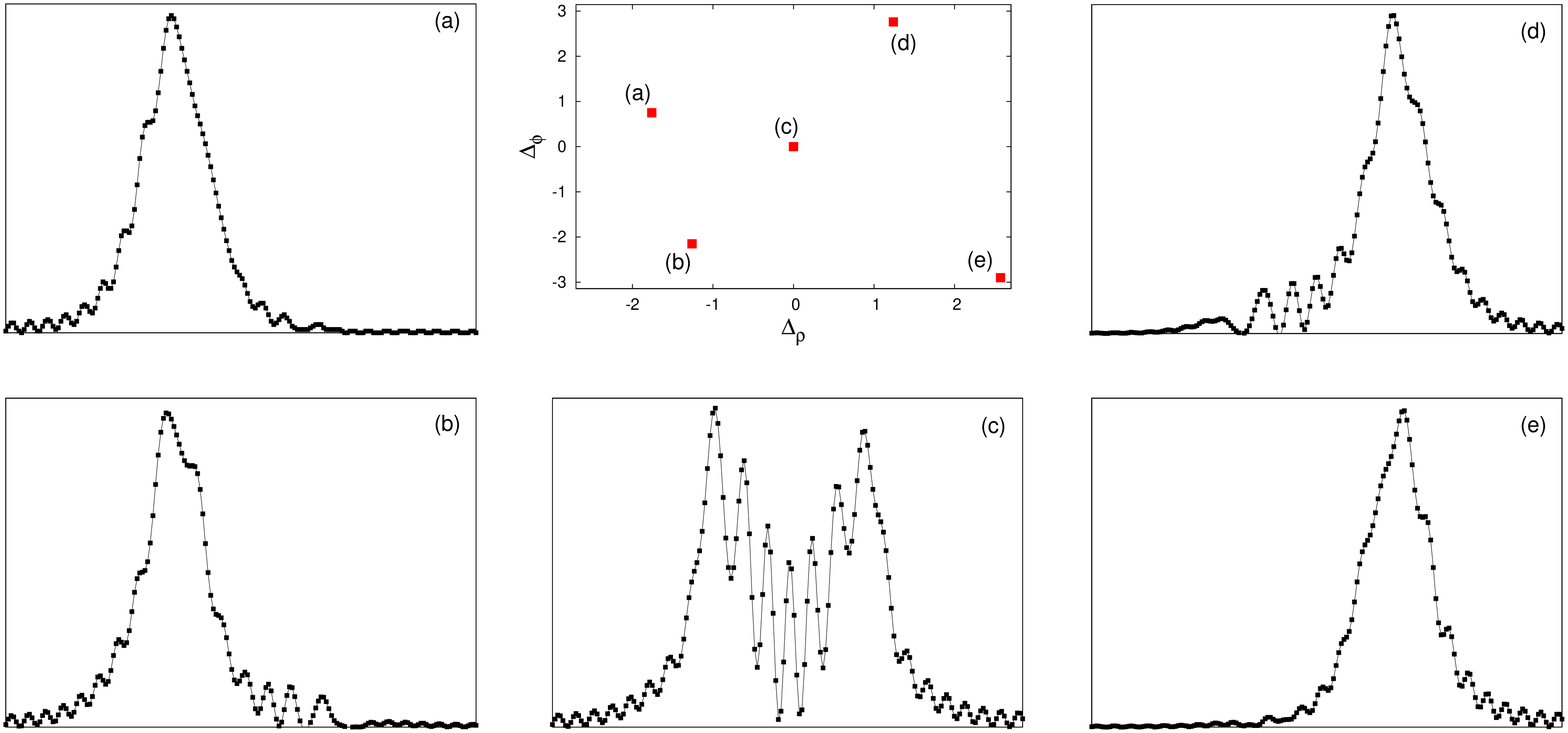}
  \caption{Five examples of different interference patterns in relation to their $(\Delta_\rho,\Delta_\phi)$-values. The case (c) is an example of a constructive interference pattern with $\Delta_\rho=\Delta_\phi=0$, 
           produced by the integrable billiard for $k=89.707$. Other cases are examples of non-constructive and/or asymmetric patterns, as either $\Delta_\rho \neq 0$ and/or $\Delta_\phi \neq 0$. They are generated by the chaotic 
           billiard, for: (a) $k=90.471$, (b) $k=92.623$, (d) $k=87.208$ and (e) $k=91.233$.} \label{fig-1Dcurrents}
\end{center} \end{figure*}

\begin{figure*}[!ht] \begin{center}
    \includegraphics[height=8.cm,width=16.cm]{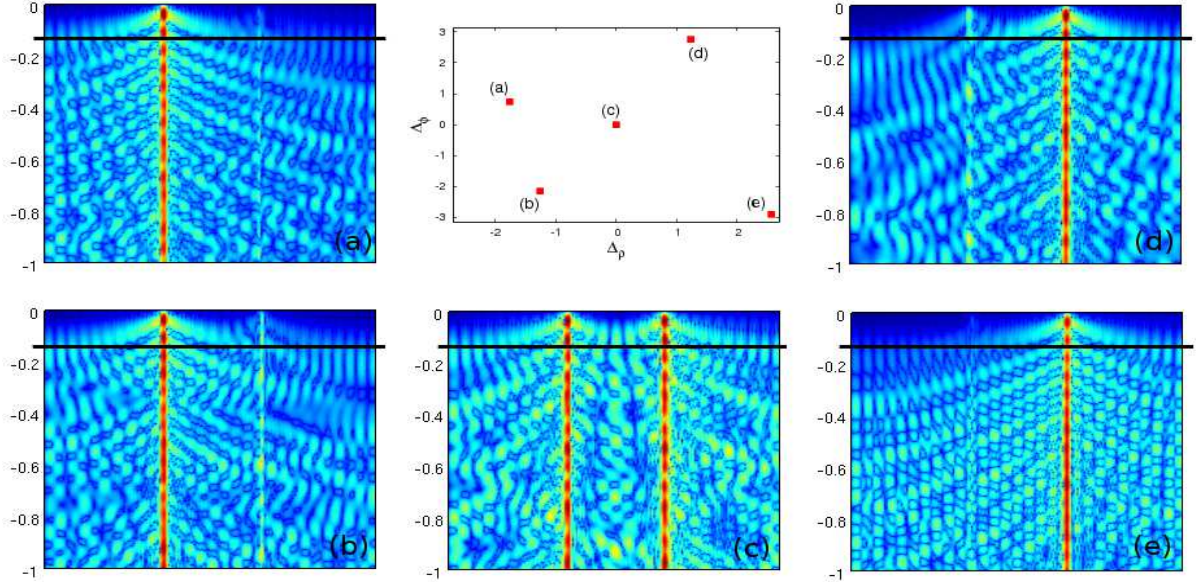}
 \caption{(color online) Two-dimensional visualization of interference of two probability currents inside the output channel. The scale indicates the intensity of the vertical component of the current 
      defined in Eq.\,(\ref{eq-interference}). Five situations correspond to the five cases of interference patterns from Fig.\,\ref{fig-1Dcurrents}. Thick line indicates the position of 
      the screen.}  \label{fig-2Dcurrents}
\end{center}  \end{figure*}

The pair of values $(\Delta_\rho,\Delta_\phi)$ captures the essential properties of an interference pattern, quantifying its discrepancy from the constructive $\Sigma$. This is illustrated in Fig.\,\ref{fig-1Dcurrents} where we show the interference patterns for five characteristic examples of $(\Delta_\rho,\Delta_\phi)$-values.
The interference pattern shown in Fig.\,\ref{fig-1Dcurrents}c having $(\Delta_\rho,\Delta_\phi)=(0,0)$ is produced by the integrable billiard, and it is essentially identical to the pattern obtained from the ideal double-slit experiment with perpendicularly impacting plane waves (with the appropriate particle's energy). Any other pattern with $(\Delta_\rho,\Delta_\phi) \neq (0,0)$ shows (at least) some departure from this ideal case. This is illustrated in Fig.\,\ref{fig-1Dcurrents}\,\,(a),(b),(d) and (e), where we show four typical examples of $\Sigma$-s generated by the chaotic billiard. For instance, the pattern in Fig.\,\ref{fig-1Dcurrents}a with $\Delta_\rho > 0$ is characterized by almost entire probability flux passing through only one slit. The pattern from Fig.\,\ref{fig-1Dcurrents}e is formed by the current exiting the billiard through one slit, in addition to a phase-shift towards the other slit ($\Delta_\rho > 0$, $\Delta_\phi < 0$). This is also clear from Fig.\,\ref{fig-2Dcurrents} where we show two-dimensional interference of probability currents in the output channel corresponding to the five situations from Fig.\,\ref{fig-1Dcurrents}. It is only in the case from Fig.\,\ref{fig-2Dcurrents}c that we have a fully symmetric two-dimensional interference characterized by both slits emitting equally strongly and in phase coherence.

The interference properties of a specific pattern $\Sigma(x)$ are basically independent of the billiard, and depend exclusively on the pattern's $(\Delta_\rho,\Delta_\phi)$-value. We verified this claim for vast range of patterns produced by different billiards with the same $(\Delta_\rho,\Delta_\phi)$-values. The pattern's properties do depend on the incoming particle's $k$-value in the sense of total number of fringes. However, since we focus only on the contrast of $\Sigma(x)$, we disregard this dependence and consider all the patterns with the same $(\Delta_\rho,\Delta_\phi)$-values to be identical (from the contrast/symmetry point of view). Although the patterns from Figs.\,\ref{fig-1Dcurrents}\,\&\,\ref{fig-2Dcurrents} were obtained from integrable and chaotic billiards for different $k$-values, they contain the properties of any billiard's $\Sigma$ with those $(\Delta_\rho,\Delta_\phi)$-values. 
In the rest of this work we will identify $\Sigma$-s only through their $(\Delta_\rho,\Delta_\phi)$-values, and in particular, distinguish only between the constructive patterns defined by $(\Delta_\rho,\Delta_\phi)=(0,0)$ and other non-constructive patterns.

\begin{figure*}[!ht] \begin{center} 
       \includegraphics[height=4.5cm,width=12.5cm]{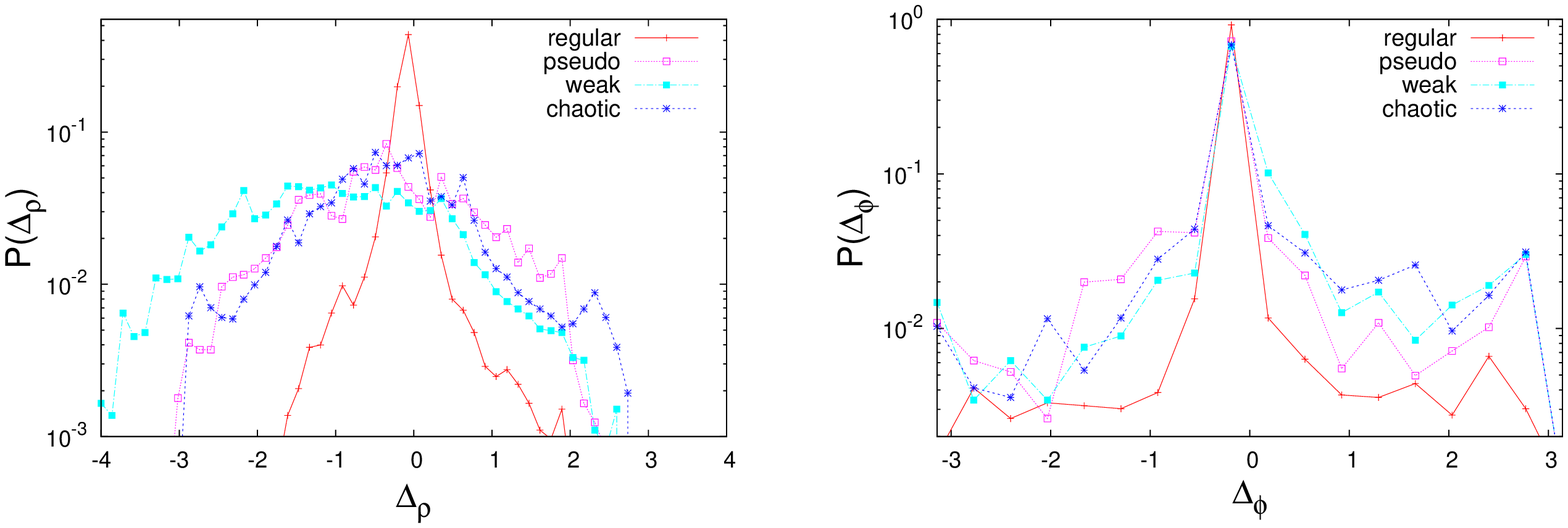}
  \caption{(color online) Data from Fig.\,\ref{fig-pdscatterplot} shown for each axis separately. Left: distributions of $\Delta_\rho$-values for all billiards (indicated by different symbols) done over 
           the entire $k$-interval from $k=85$ to $k=94$. Right: the same for $\Delta_\phi$-values. } \label{fig-pd-distributions} 
\end{center} \end{figure*}

\paragraph{The distribution of interference patterns.} The remaining question we answer in the rest of this Section refers to the typical proportion of constructive versus non-constructive patterns, in relation to four billiards. After establishing a connection between a pattern's constructiveness properties and its $(\Delta_\rho,\Delta_\phi)$-value, we statistically study the distributions of $\Delta_\rho$ and $\Delta_\phi$. In Fig.\,\ref{fig-pdscatterplot} we show two-dimensional scatter plot distribution of $\Delta_\rho$-$\Delta_\phi$, done over the entire examined $k$-interval. In the integrable case the patterns are visibly concentrated around a single central peak at $(\Delta_\rho,\Delta_\phi)=(0,0)$. In opposition to this, non-integrable billiards display rather diffused and more uniform distributions, without a clear central peak.
While in the integrable case the constructive patterns are overwhelmingly pre\-sent over a large $k$-interval, other billiards exhibit more significant fractions of non-constructive patterns with large  $|\Delta_\rho|$,
but also larger fraction of asymmetric patterns with large $|\Delta_\phi|$. We illustrate this further by showing in Fig.\,\ref{fig-pd-distributions} projected (marginal) distributions of $\Delta_\rho$ and $\Delta_\phi$, simultaneously comparing all the four billiards.
The distributions of $\Delta_\rho$-values, $P(\Delta_\rho)$, shown on the left reports a clear discrepancy between the integrable billiard and the non-integrable billiards. $P(\Delta_\rho)$ in the integrable case exhibits a clear peak which exponentially decays away from $\Delta_\rho=0$. On the other hand, $P(\Delta_\rho)$ for other billiards show a gaussian-like profile with much larger dispersion.
The distributions of phase-shifts $P(\Delta_\phi)$ display less prominent discrepancy among the billiards, although the same trend is visible. While statistics for both $\Delta_\rho$ and $\Delta_\phi$ show a distinction between integrable billiard and non-integrable billiards, the difference among the non-integrable billiards themselves is much less visible.


\section{Discussion and Conclusions}  \label{DiscussionConclusions}

We carried out a computational experiment, designed as a model of scattering of quantum particles through billiards with variable geometries, i.e. variable classical integrability properties. In particular, we considered a fully integrable regular triangular billiard, and three non-integrable billiards: triangular pseudo-integrable, triangular weak-mixing, and Sinai-type strongly chaotic (K-system). Billiards were designed with a double-slit opening on their base in a way to allow for two scattered probability currents to interfere and create an interference pattern observable on a screen. The properties of the interference patterns were investigated in relation to the classical integrability of the billiards. Experiments were done for a range of particles' energies, and the interference properties of the obtained patterns were studied statistically, comparing among the billiards.

Dramatic difference between the interference patterns exhibited by the integrable (regular) billiard and non-integrable billiards was found. In the integrable case an overwhelming fraction of the 
observed patterns is constructive, i.e. resembling the outcome of the ideal plane-wave double-slit experiment. In contrast to this, non-integrable billiards display a vast variety of interference patterns, with a large fraction of them being asymmetric or of low visibility (contrast ratio). We substantiated our study by introducing the quantities $\Delta_\rho$ and $\Delta_\phi$ in Eq.\,(\ref{eq-pd-definition}) that quantify the phase-symmetry and intensity-symmetry between the two slits. As clear from Fig.\,\ref{fig-pdscatterplot}, the integrable billiard exhibits a sharp peak in the joint distribution of these parameters at $(\Delta_\rho,\Delta_\phi)=(0,0)$, whereas non-integrable billiards show much more dispersed distribution of $(\Delta_\rho,\Delta_\phi)$.

Although all the experiments were conducted with a fixed height of the potential barrier, we investigated the effect of changing its height. Raising the barrier makes the resonant peaks (cf. Fig.\,\ref{fig-resonances}) narrower and gives a better clarity of the interference pattern, as this way the particle ``feels'' the billiard's geometry better. However, this also reduces the total output flux for all $k$-values, making many interference patterns unusable for statistical study. Our selection of the barrier height optimizes between these two issues. 

The interference patterns change smoothly with $k$. The fact that non-integrable billiards display a larger variety of different patterns, also testifies about their more rapid change with increasing $k$. Additional insights into the mechanism of quantum dephasing might be obtained by studying the speed of the structural change of the interference patterns with $k$.

Along the lines of conclusions reached in \cite{casatiprosen}, we thus re-confirmed the connection between the integrability of a classical system, and the appearance of quantum dephasing responsible for the destruction of the interference patterns. The classically chaotic geometry seems to be able to generate dephasing in both time-domain (as shown in \cite{casatiprosen}) and in energy-domain. Interestingly, while the distinction between integrable and non-integrable billiards was clearly revealed, the difference among the non-integrable billiards is much less visible. As it appears from Figs.\,\ref{fig-pdscatterplot}\,\&\,\ref{fig-pd-distributions}, the presence of non-constructive interference patterns is similarly prominent for all three non-integrable billiards. It seems that all three of them are equally able to generate dephasing, and no clear conclusion can be made from these results regarding the strength of the dephasing induced by billiards of different non-integrability levels. Further work and possibly new experiments are needed to clarify this interesting question.

The computational experiment presented here can be easily constructed and carried out in a real laboratory.\\[0.3cm]


\noindent {\bf Acknowledgments.} This work was supported the DFG through the project FOR868 and by the National Program P1-0044 (Slovenia). ZL is very grateful to J. Stefan Institute and University of Ljubljana (Slovenia) where a large part of this work was done. Useful comments and suggestions by prof.s B. Tadi\'c and A. Pikovsky are acknowledged. Thanks to M. \v Suvakov, G. Veble, I. Pi\v zorn and M. Horvat for constructive discussions. Special thanks to R. Krivec for maintenance of the computing resources where most of the numerics was done.

\end{document}